# Crowdfunding Public Projects:

## Collaborative Governance for Achieving Citizen Co-funding of Public Goods

Sounman Hong

Yonsei University

Jungmin Ryu

Yonsei University

*Abstract*

This study explores the potential of crowdfunding as a tool for achieving "citizen co-funding" of public projects. Focusing on philanthropic crowdfunding, we examine whether collaborative projects between public and private organizations are more successful in fundraising than projects initiated solely by private organizations. We argue that government involvement in crowdfunding provides some type of accreditation or certification that attests to a project's aim to achieve public rather than private goals, thereby mitigating information asymmetry and improving mutual trust between creators (i.e., private sector organizations) and funders (i.e., crowd). To support this argument, we show that crowdfunding projects with government involvement achieved a greater success rate and attracted a greater amount of funding than comparable projects without government involvement. This evidence shows that governments may take advantage of crowdfunding to "co-fund" public projects with the citizenry for addressing the complex challenges that we face in the twenty-first century.

# 1. INTRODUCTION

Crowdfunding is an increasingly popular and widespread form of online fundraising, whereby small individual contributions made through an online platform by the crowd are pooled to fund a project and achieve a particular goal. Research has reported that a rapidly growing volume of money is collected through crowdfunding in many parts of the world (Agrawal, Catalini, & Goldfarb, 2014; Belleflamme, Lambert, & Schwienbacher, 2014; Burtch, Ghose, & Wattal, 2013; Mollick, 2014; Mollick & Nanda, 2015; Mayer, 2016, 2018; Hong & Ryu, 2018). For instance, according to a report by Massolution, a research firm analyzing crowdfunding market trends, the worldwide fundraising volume through crowdfunding was estimated to be 34 billion USD in 2015, a more than tenfold increase from 2.7 billion USD in 2012.

Somewhat puzzling evidence indicates that many crowdfunding projects are successful in fundraising, even if their expected returns are very low or even negative (Boudreau, Jeppesen, Reichstein, & Rullani, 2015; Evans, 2015; Griffin, 2012; Hazen, 2011). That is, the reward structure of many crowdfunding projects closely resembles charitable donations; people make voluntary contributions to public goods through crowdfunding platforms (Boudreau et al., 2015). In order to explain this conundrum, many researchers have proposed that crowdfunders may enjoy some intangible rewards from their participation (Burtch et al., 2013). Such an observation is not limited to philanthropic crowdfunding but applies more generally to a much wider class of projects, including entrepreneurial ones (Boudreau et al., 2015).

The evidence that crowdfunding resembles charitable donations suggests that it could potentially be used to address not only private but also public missions. Recognizing this potential, governments in many parts of the world have started to actively engage in partnering with non-profit sectors to initiate projects through crowdfunding platforms (Lee, Zhao, &

Hassna, 2016; Miglietta, Parisi, Pessione, & Servato, 2014; Hong & Ryu, 2018). However, despite the growing participation of governments in the crowdfunding industry, little is known about whether and how they may take advantage of this innovation to achieve public goals. The lack of scholarly attention to public-private partnerships in crowdfunding is an important omission in the literature, given that many governments around the world are already actively utilizing it. For instance, in Korea, where the evidence of this study comes from, a large number of local and central government agencies have started to actively use crowdfunding platforms to fund various projects that could potentially contribute to public missions. Commentators have applauded this trend as a remarkable innovation in the way in which governments operate. However, to our best knowledge, no empirical evidence exists to support this claim.

In this study, we follow previous research (e.g., de Vries, Hanna, Bekkers, & Tummers, 2016) and define "public sector innovation" generally as public organizations' adoption of "an idea, practice, or object that is perceived as new by an individual or other unit of adoption" (Rogers, 2003, 12). De Vries et al. (2016) proposed four specific types of innovation: (1) process innovation, (2) product or service innovation, (3) governance innovation, and (4) conceptual innovation. Among the four types, public-sector crowdfunding may be classified as an example of both process and governance innovations as it offers a new process as well as a new form of governance for addressing societal problems. Crowdfunding offers a new process in which governments can solicit creative solutions from nonprofit organizations (NPOs) for a given social problem, and collect contributions from a large number of funders more efficiently than before.[1] It also enables a new form of collaborative governance in which stakeholders pool their innovation assets to develop joint solutions for social problems. Specifically, the government oversees the whole process while steering the overall direction of social investments and signaling that a project aims to achieve social rather than private goals. NPOs use their expertise to propose creative and specific solutions for solving the social problems set forth by the government. Funders evaluate the proposed solutions by contributing to the projects they value.

---

[1] The so-called sharing economy generally involves peer-to-peer-based activities of sharing the access to goods or services, as coordinated through an online platform (Hong & Lee, 2018a, 2018b). Crowdfunding is often classified as a specific type of sharing economy as it involves peer-to peer financing (Hamari, Sjöklint & Ukkonen, 2016).

Governments' use of crowdfunding may be regarded as a public sector innovation with normative and economic benefits. First, crowdfunding may be reconcilable with democratic values as it "democratizes" financing. The amount of money raised by crowdfunding may serve as a credible proxy for the collective demand (i.e., needs) of the citizenry for the projects; crowdfunding allows citizens to "vote with their dollars" online to bring ideas into reality. Second, crowdfunding may especially be useful for governments to carry forward politically contentious public services. Taxpayers would be against the idea that their money is spent on projects that do not fit with their political views, but they may care less if the money is raised from the voluntary contributions of crowdfunders. That is, with crowdfunding, governments may be shielded from criticisms that they pursue projects to serve their own benefits or political ideology, rather than achieve broader public missions. Third, crowdfunding may also allow for the funding of public projects without distorting the choices of those who pay for the projects. In contrast, taxation has generally negative impacts on social welfare as it distorts taxpayers' choices; for instance, if the wage tax rate increases, then laborers decide to work less. Such a welfare loss does not accrue in crowdfunded public projects, as citizens' voluntarily donations do not distort their own economic activities.

In this study, we evaluate the potential of crowdfunding as a tool for achieving innovation in the public sector. We argue that governments may use crowdfunding to collaborate with the private sector to achieve a public mission. That is, a crowdfunding platform may function as an online infrastructure where a public-private sector collaboration or partnership may be formed to address the complex challenges that we face in the twenty-first century. To test this argument, we focus on philanthropic crowdfunding and examine whether collaborative projects of public and private organizations are more successful in fundraising than projects initiated solely by private organizations. This empirical investigation aims to establish whether government involvement significantly improves the capability of private sector organizations (e.g., philanthropic organizations) in soliciting citizen contributions to public goals, in which case crowdfunding may be used to solve various collective action problems.

Overall, findings suggest that, with all else being equal, government involvement has significant positive impacts on crowdfunding performance, as measured by success rate and funding amount. To explain this finding, we hypothesize that the participation of government agencies could improve crowdfunding performance by mitigating the information asymmetry between the creator (i.e., the private sector organization) and funders (i.e., the crowd). Specifically, we argue that government involvement provides some type of accreditation or certification that attests that crowdfunding projects truly aim to achieve public rather than private goals, ultimately improving citizens' trust in the projects. In what follows, we show evidence that supports this hypothesis.

## 2. PUBLIC SERVICE INNOVATION AND CROWDFUNDING

In many parts of the world, public sector organizations are faced with the growing needs of social programs to address health, inequality, and aging issues, while at the same time experiencing severe resource constraints. In response, there is considerable interest in how governments may become more innovative in addressing such challenges (Arundel, Casali, &

Hollanders, 2015; Borins, 2008; Osborne & Brown, 2011; Hartley, Sørensen, & Torfing, 2013; Bugge & Bloch, 2016; Demircioglu, 2017; Demircioglu & Audretsch, 2017). However, despite such interest, prior research has focused primarily on how private sector organizations innovate but have paid little attention to the public sector context. This paucity of research is surprising, given that innovation in the public sector generally has significant spill-over; its benefits reach far beyond the public sector's boundaries (e.g., Julnes & Gibson, 2015; Torfing & Triantafillou, 2016).

Prior research has identified several broad strategies that can be implemented by public sector organizations to foster innovation (Hartley et al., 2013). Among them, this study focuses on the strategies that rely on the idea that collaborative governance involving both public and private actors can prompt and sustain public sector innovation. This type of strategy is often called "collaborative innovation." The rationale underlying collaborative innovation mainly stems from theories of network governance and organizational learning that emphasize how collaborative processes involving a broad range of actors are conducive to finding innovative solutions to complex problems (e.g., Engestrom, 2008; Koppenjan & Klijn, 2004; Clark, Brudney, & Jang, 2013; Collm & Schedler, 2014; Scupola & Zanfei 2016; Gascó, 2017; Lindsay et al. 2017). The primary driving force of collaborative innovation may be that such participatory governance opens up the innovation processes to a wide range of stakeholders facing common problems (Bommert, 2010; Hartley et al., 2013; Lindsay et al. 2017). Those stakeholders, having a comparative advantage in different innovation assets, are integrated into the innovation cycle to pool their assets and develop joint solutions (Bommert, 2010; Hartley et al., 2013).

Although scholars generally agree with the benefits of collaborative approaches in innovation, they also point out the difficulties in forming a well-functioning partnership for collaboration. Prior research identified the lack of trust among participating actors as a key barrier against collaboration (e.g., Agranoff & McGuire, 2001; Thomas, 2003; Huxham & Vangen, 2005; Bryson, Crosby, & Stone, 2006, 2015; Thomson, Perry, & Miller, 2008; Hong & Kim, 2018; see also Kelman, Hong, & Turbitt, 2012). In fact, collaborative governance may not be sustainable if participating organizations are concerned about the opportunism (more formally "moral hazard") of other participants. Moral hazard is a significant concern, especially when asymmetric information exists among participating organizations (Hölmstrom, 1979). Therefore, the formation of a successful collaborative partnership rests on whether and how information is shared among partners, at least before mutual trust is formed. As building trust among participants is not always feasible (e.g., too time-consuming), maintaining a high level of information transparency is crucial in expanding the network of any collaborative governance (e.g., Bellamy, Raab, Warren, & Heeney, 2007).

The recent development of information technology and online infrastructure, including crowdfunding platforms, has transformed the landscape of collaborative governance and innovation. For instance, crowdfunding provides an unprecedented potential for any person or organization with innovative ideas to overcome their resource constraints through small contributions by the public. However, it should be noted that information asymmetry may be especially problematic in crowdfunding due to the large number of investors (Strausz, 2017). Due to asymmetric information, funders may not reveal their true preferences, whereas creators may embezzle the collected fund. This is where the role of governments comes in. In this study, we show that the government participation in crowdfunding projects contributes to the formation of collaboration by mitigating the information asymmetry between the creator and funders.

## 3. BACKGROUND: GOVERNMENT INVOLVEMENT IN CROWDFUNDING

This study's evidence is based on observations from a major Korean crowdfunding platform, Wadiz, and was collected from March to December 2016. A Korean platform was selected as a case study because of the Korean government's active involvement in the crowdfunding industry. The so-called "Crowdfunding Law" [2] was approved by the Korean National Assembly in July 2015 and took effect in January 2016. This law terminated the severe financial regulations that had been imposed on crowdfunding; as a result, the crowdfunding industry came to face a lighter regulatory burden than other traditional financial industries. Subsequently, the Korean government announced the "Crowdfunding Promotion Plan" in January 2016 to encourage collaborations among various public authorities (i.e., both central and local government agencies) and private sector organizations for crowdfunding to promote entrepreneurship and economy. These endeavors produced a substantial number of cases in which government agencies and NPOs formed collaborative partnerships through crowdfunding platforms.

Wadiz was founded in May 2012 by a group of entrepreneurs and was officially approved by the Korean government as a crowdfunding platform in January 2016. As of December 2017, Wadiz is by far the largest crowdfunding platform in Korea, with a cumulative fundraising amount of 44.4 billion Korean Won (which translates into about 40 million USD). On this platform, there are two types of crowdfunding project models: equity-based and reward-based models. The equity-based models are similar to angel investments, in which the funders become shareholders who receive financial rewards based on the realized profits. On the other hand, the reward-based models provide only nonfinancial rewards to the funders. Many projects give funders some small souvenirs; for instance, an NPO that aims to support people who are blind or with low vision gave a braille bracelet to funders. As the reward-based model does not provide any substantial amount of financial rewards in exchange for contributions, these projects may be viewed as charitable donation-based crowdfunding, commonly observed in major international platforms such as Kickstarter. The reward-based models are further classified into eleven categories including technology, fashion and beauty, food, design, online cartoons, video games, book publishing, arts, travel, sports, and public projects. The sample used in this study includes all crowdfunding projects that are classified as "public projects" and that appeared on the site from March to December 2016.

---

[2] Amendment to the "Financial Investment Services and Capital Markets Act."

The roles of government agencies in crowdfunding projects vary to some extent, but the government's primary function is to determine areas for their involvement and perform "screening" (i.e., the selection of projects that are feasible and may contribute to public missions). Specifically, the government first announces some broad areas of support and involvement (e.g., promotion of well-being in rural regions). They then solicit creative solutions from private sector organizations (i.e., NPOs) to address the social problems related to the announced areas; such solicitations may be viewed as the crowdsourcing of ideas reported in previous literature (Linders, 2012; Mergel & Desouza, 2013; Maheshwari & Janssen, 2014; Mergel, 2015a, 2015b; Sieber & Johnson, 2015; Moon, 2018). Next, the government conducts reviews of crowdfunding proposals submitted by NPOs and individuals, and selects those projects that are feasible, fit the announced areas of involvement, and have significant potential to contribute to public missions if properly funded and carried forward. The selected projects are delivered to the crowdfunding platform and identified as such. For instance, Wadiz shows a message such as "this project is pursued as a collaboration with [the name of public agencies]." As explained above, such government involvement is mainly pursued in "public projects," as classified by the platform.

Governments' involvement in crowdfunding projects may be viewed as a case of collaborative innovation for the following reasons. First, crowdfunding public projects allow governments, NPOs, and funders to focus on their own comparative advantages; each party brings distinct capabilities to create potentially greater synergy in the collaboration (Bryson et al., 2006; Austin & Seitanidi, 2012; Hong & Kim, 2018). The government provides the accreditation that a project aims to achieve social rather than private goals (Rivenbark & Menter, 2006). NPOs may reduce their fundraising costs and rather focus on their key areas of expertise (i.e., management of the project). Funders may be more willing to share their resources as they become less concerned about potential embezzlement. Second, although the involved stakeholders seem to act independently, they work to achieve a common public goal. That is, the crowdfunding platform enables stakeholders to pool their innovation assets for the development of *joint solution*s. Specifically, the government steers the direction of social investments by identifying areas that need more care and also monitors the opportunism of nonprofit partners. NPOs implement and manage the projects, while funders provide voluntary contributions based on their evaluations of the projects' social values. Here, although all the stakeholders may not necessarily participate in mutual decision-making as is common in many traditional forms of collaborative governance, the crowdsourcing platform allows them to produce a joint solution for achieving a common goal.

## 4. HYPOTHESES

*4.1 Information Asymmetry and Government Participation*

Previous research in corporate finance has well reported how the existence of information asymmetry, a situation where the firm has more or better information than investors, may present a barrier to efficient financing and investment decisions (Myers & Majluf, 1984). This insight extends to the case in which an NPO attracts contributions from the citizenry. Citizens have little information about how the donation will be spent once the contribution is made; it may be spent to address social problems as announced by the NPO, but it may also be used for the organization's own interests—a problem termed as "moral hazard." In the context of crowdfunding, there may exist information asymmetry between the creator (i.e., the NPO) and the funder (i.e., the crowd), which may lead to a potential moral hazard for the creator.

Information asymmetry may distort the outcome of crowdfunding in two different ways. First, the creator may embezzle some of the collected finance, which represents the abovementioned moral hazard. Second, expecting the risks of a moral hazard, the crowd may abstain from contributing. That is, even those who fully sympathize with and support the missions of the nonprofit organization may be unwilling to contribute due to the uncertainty involved in how the contribution will be spent. In sum, the information asymmetry problem has a distortive effect on the choices of both the creator and the funders. In what follows, this study focuses on the latter (i.e., the crowd's abstention from contribution due to information asymmetry) and shows how a public-private partnership in crowdfunding may contribute to reducing this distortion.[3]

To be more specific, this information asymmetry problem may be mitigated under certain conditions. For instance, the distortion will be less significant if the creator has a good reputation and is trusted by the public (Van Slyke, 2006). However, we focus in this study on two variables among the many possible factors: (1) the creator's effort to maintain a high level of transparency, and (2) government involvement. We hypothesize that these two factors may significantly reduce information asymmetry in crowdfunding projects.

---

[3] Government involvement may also reduce the transaction costs of the crowdfunding process. For instance, without government support, private sector organizations may have to work to improve the transparency of their activities, which may be costly. Government involvement thus reduces the need for these activities, which may be seen as resulting in a decrease in the transaction costs of the crowdfunding process.

*4.2 Government Participation and Crowd Fundraising*

Prior research has emphasized the critical importance of transparency for the successful management of NPOs. Specifically, the voluntary disclosure of information by NPOs may have a significantly positive impact on the amount of donations by improving donor perceptions (Gandía, 2011). Further, the availability of the Internet is found to be conducive to an environment in which NPOs can more effectively achieve a high level of transparency (Gandía, 2011).

The same logic may extend to the case where NPOs reach out to crowdfunding platforms in order to expand their donor base. Previous research has stressed the importance of maintaining transparency for the successful performance of crowdfunding activities (e.g., Carvajal, García-Avilés, & González, 2012; Gerber & Hui, 2013). In fact, a greater level of transparency achieved by the creator's voluntary information disclosure may promote crowdfunders' trust in the project, resulting in an increase in the amount of contributions. In other words, with a greater level of transparency, the public may become less worried about the potential moral hazard of the NPOs, and thus may more willingly participate in the project. Therefore, we have the following hypothesis:

> *H1*: A greater level of transparency achieved by the creator's voluntary information disclosure may improve crowdfunders' trust, thereby resulting in an increase in contributions.

To be clear, this benefit of transparency stems from the distrust between the creator and funders; the information asymmetry creates a concern for funders about the creator's potential moral hazard behaviors (e.g., embezzlement of the contributions). Besides the creator's effort for information disclosure, we argue that governments' participation in, or support of, crowdfunding projects may also function as an instrument for improving the funders' trust, thereby leading to an increase in the amount of contributions. In the case of government participation, governments and NPOs jointly create a crowdfunding project; in contrast, government support may take a variety of different modes, including government subsidy (provided as matching funds to crowdfunding, for instance), a review of the project's feasibility and social desirability, ex-post investigation of the creator's moral hazard, and so on. Thus, we argue that government participation or support of crowdfunding projects will also result in an increase in the amount of contributions. In this vein, the following hypothesis is proposed;

> *H2*: Government participation or support of crowdfunding projects may improve crowdfunders' trust, thereby resulting in an increase in contributions.

The above hypothesis may be generalizable to a wide variety of forms of government involvement. However, among the many possible forms, this study explores the effect of arguably the weakest form of government involvement: a government agency conducts a review of the ideas underlying crowdfunding projects in terms of their feasibility and social desirability and provides certification that the projects are in line with public missions.

We have thus far explained that both the creator's effort to improve transparency and government support may have positive effects on the performance of crowd-based financing. An important insight is that both variables mitigate information asymmetry and improve the crowd's trust. The fact that improved transparency and government support both operate through the same channel (i.e., improvement in citizen trust) may indicate that they are *substitutes*. Without any government involvement, the creator's effort to improve transparency may be critical in the success of crowd fundraising. On the other hand, with government participation or support, transparency may have a smaller impact. This "substitution hypothesis" indicates that government participation or support would *negatively* moderate the association between transparency and successful crowd fundraising.

> *H3*: Government participation or support of crowdfunding projects may work as substitutes for the creator's effort to improve transparency ("substitution hypothesis")

*4.3 Government Participation and Moral Hazard*

This research has proposed that, with government involvement, the crowd will be less concerned about the potential moral hazard of the creator. To be clear, however, we do not argue that the actual level of moral hazard (e.g., the creator's embezzlement) would decrease with government participation or support. Of course, the level of moral hazard may decrease if government participation involves rigorous audits on how the contributed money is spent, but it is not clear as to whether the same will happen with other forms of government involvement (e.g., the government reviewing the feasibility and social desirability of projects). Although this issue may be related to the hypotheses that we formulated, the constructed dataset does not allow us to investigate this postulation. We thus leave this topic of research unexplored for future research.

*4.4 Summary: Theoretical Relationship*

The hypothesized relationships among government involvement, transparency, and crowdfunding performance are visualized in Figure 1. The figure shows that the transparency of crowdfunding projects has a positive association with crowdfunding performance (hypothesis 1). Further, it is also clear from the figure that government involvement has a positive impact on crowdfunding performance holding the transparency level constant (hypothesis 2). Lastly, the figure describes the diminishing "public participation premium" in crowdfunding as the projects become increasingly transparent; the hypothesis is that government involvement and transparency work as if they were substitutes.

## 5. METHODS AND DATA

*5.1 Method*

To evaluate the hypotheses proposed above, we collected data from a Korean crowdfunding platform from March to December 2016. We used the individual crowdfunding projects listed on the crowdfunding platform as the units of analysis. Our examination was to verify whether the crowdfunding projects with a higher level of transparency and/or government support were more successful, with all else being equal. Specifically, our model is as follows:

$$Y_i = \alpha + \rho T_i + \tau P_i + X_i \gamma + \lambda_i + \nu_i \qquad (1)$$

where $Y_i$ is the dependent variable (i.e., the crowdfunding performance) for a crowdfunding project *i*. We used two different indicators of performance: (1) the project success rate, as measured by the amount of contributions raised, divided by the project's funding goal, and (2) the total amount of contributions raised through crowdfunding. Both were log-transformed to enable the interpretation of coefficients as percentage impacts. On the other hand, $T_i$ and $P_i$ are the two treatment variables, the transparency level and an indicator of government support, respectively. Thus, the coefficients $\rho$ and $\tau$ are expected to both be positive, according to hypotheses 1 and 2. The vector $X_i$ comprises a set of covariates, including the number of days the project was listed on the platform, an indicator of a "keep-it-all" type, and the funding goal of the project. $\lambda_i$ indicates a list of dummy variables for various types of potential beneficiaries of the project, whereas $v_i$ is the error term.

We also estimated the following equation to investigate the "substitution hypothesis," namely whether the level of transparency matters less if the project is supported by governments, as the crowd will trust in the project even with a lack of transparency. In other words, we tested whether government support of a crowdfunding project mitigates the information asymmetry problem, thereby facilitating the outcomes of crowdfunding. We tested this argument with the following model:

$$Y_i = \alpha + \rho T_i + \tau P_i + \theta T_i P_i + X_i \gamma + \lambda_i + v_i \qquad (2)$$

where the coefficient of the interaction term $\theta$ is of primary interest. We expected that the coefficient $\theta$ would be negative. Thus, the benefit of the creator's efforts for maintaining transparency decreases when the crowdfunding project is supported by government agencies. In equation (2), the definitions of all the remaining variables are the same as in equation (1).

*5.2 Data*

The two dependent variables of this study are crowdfunding success rate and funding amount, whereas the two treatment variables comprise an indicator of government involvement and the transparency level of a project. The government involvement indicator is a dummy variable, with the value 1 if the project is supported by a government agency and 0 otherwise. This variable is easily identifiable as projects that received government support are announced by the platform, for instance, by "this project is pursued as a collaboration with [the name of a public organization]."

On the other hand, the other treatment variable, the transparency level, is a weighted average of three measures: (1) the number of communications between the creator and those visiting the site of the project, (2) the number of updates the creator posted, and (3) the measured level of detail of the plan on how the raised funds will be spent. Here, a "communication" is the comment the creator posted in response to questions posted on the platform (i.e., two-way communication), whereas an "update" is what the creator posted about the project's progress (i.e., one-way communication). We also evaluated the creator's spending plan for the raised funds. Some projects simply stated that, for example, "the raised fund will be used to support children with blood cancer," while others had a much more detailed plan with information on the schedule, process, or specific beneficiaries. These three variables are standardized to have a zero mean and one standard deviation and are then averaged to yield the transparency variable.

In addition, we collected a number of covariates. To begin with, we controlled for the number of days the project was listed on the platform (i.e., duration), as the crowdfunding outcomes may depend on how long funding had been solicited. Second, we also included an indicator of a "keep-it-all" type. In crowdfunding projects, a "keep-it-all" type involves a creator who keeps the entire amount raised, regardless of whether the outcome met their funding goal, whereas an "all-or-nothing" type involves a creator who keeps nothing if the goal is not achieved. Third, we controlled for the funding goal in models, with the amount of funding as the dependent variable. Fourth, we included a set of dummy variables indicating various types of potential beneficiaries of the project. The beneficiaries are categorized as the disabled, women, families in poverty, the elderly, multicultural families, foreign aid, cultural issues, and others.[4] We decided to control for indicators of these categories, as crowdfunding outcomes may vary across them. Table 1 lists the summary statistics for all variables used in the analysis. All data were collected from Wadiz.

In this study, we analyzed 110 projects listed on the Wadiz platform as "public projects" from March to December 2016. Out of these 110 cases, 37 were pursued as collaborations between a governmental agency and private sector organizations, whereas the remaining 73 were carried out solely by NPOs. The average funding goal was about 2.7 million KRW (which translates to about 2,500 USD per project), ranging approximately from 500 to 20,000 USD. Out of the 110 projects, foreign aid projects constitute the largest group in terms of intended beneficiaries with 16 projects, followed by cases that aim to support the elderly (13 projects), families in poverty (nine projects), cultural issues (nine projects), and the disabled (eight projects). On average, funding had been solicited for 29.4 days on the platform with a range of as short as ten days to as long as 105 days.

## 6. RESULTS

The results of our analyses are presented in Tables 2 and 3. In each of the tables, we used a different measure of crowdfunding performance as a dependent variable. In Table 2, the dependent variable is the crowdfunding success rate, as measured by the amount funded divided by the funding goal. To check the robustness of our finding, we repeated the same analysis in Table 3, with the total amount raised as the dependent variable.

In Table 2, we demonstrate our results in four different models. All the presented models control for the covariates, including the set of dummy variables that indicate the project beneficiaries. In columns 1 to 3, we show that both government involvement and transparency have positive impacts on the project success rate. Specifically, findings suggest that the existence of government support is associated with an approximately 64% increase in the crowdfunding success rate (column 3 of Table 2). On the other hand, a one standard deviation increase in transparency is found to be associated with an approximately 41% increase in success rate (column 3 of Table 2). Overall, the findings provide support for hypotheses 1 and 2.

---

[4] The group "others" refers to a project for which the beneficiary is not clear. For example, a project claimed that the amount raised will be spent to solve a dispute between Korea and Japan over the sovereignty of Dokdo island. Such a project is coded as "others" because beneficiaries are not specifically identified.

In column 4 of Table 2, we present our results with an addition of the interaction term between transparency and government support. As can be seen, the coefficient of the interaction term is significantly negative as hypothesized, which indicates that transparency and government involvement substitute each other to some extent; that is, the provision of government support is expected to yield a greater reward when the project is less transparent. Specifically, findings indicate that, in the absence of government support, a one standard deviation increase in transparency is associated with a success rate increase of about 72%. However, with government support, the same increase in transparency is associated with a mere 23% increase in the performance measure.

We confirmed the robustness of our findings in Table 3 by using an alternative measure of crowdfunding performance (i.e., total amount funded) as the dependent variable. In Table 3, we also included an additional variable (i.e., the funding goal) as a control variable, because the total amount raised may be significantly influenced by the targeted goal as set out by the creator. All the models in Table 3 control for the covariates, including the set of dummy variables, indicating the beneficiaries of the project as well as the funding goal. Since we log-transformed the values of the two dependent variables in Tables 2 and 3, we can interpret the coefficients as percentage changes, enabling us to directly compare the two sets of estimates. Overall, the results in Table 3 are largely consistent with what we found in Table 2.

In column 4 of Table 3, we report our results with the interaction term between transparency and government support. Here again, the coefficients are very similar to what we found in column 4 of Table 2, providing support for the "substitution hypothesis." Specifically, findings indicate that, in the absence of government support, a one standard deviation increase in transparency is associated with an approximately 75% increase in success rate. However, with government support, the same increase in transparency is associated with only a 28% increase in the performance measure. Overall, the results once again indicate that collaborative governance between public and private sector organizations yield a greater benefit when the project lacks transparency.

Figure 2 shows the estimates from Table 2 graphically to refocus attention on the theoretical claim. The figure shows the relationship among government involvement, transparency of information, and our empirical measure of crowdfunding performance (i.e., the log-transformed success rate). We calculated the values in the figure directly from the coefficients in column 4 of Table 2. To do so, we set covariates at their mean values, took the product of the coefficients and values, and then summed up these quantities (3.78). We then added this value to those of different configurations of the key variables (i.e., transparency, government support, and the interaction between the two). In the figure, "high transparency" and "low transparency" are transparency variables that represent the mean plus one standard deviation and the mean minus one standard deviation, respectively. Figure 2 closely resembles the theoretical expectations appearing in Figure 1, providing empirical support for our hypotheses.

# 7. THE FUTURE OF PUBLIC SECTOR CROWDFUNDING

In this study, we focused on highlighting the positive potential of governments' use of crowdfunding. Like other modes of public sector innovation, however, public sector crowdfunding may also involve some potential risks. In this section, we propose a number of potentially negative aspects of public sector crowdfunding and discuss their implications for social welfare and public policymaking. Although we are not in a position to evaluate the full impact of public sector crowdfunding, given that the new technology is still in its infancy, we believe it is meaningful to consider the prospect of its wider consequences.

First, a critical concern may be that the rise of public sector crowdfunding may incentivize governments to pursue "hidden privatization"; governments downsize their roles and instead use the new technology to outsource public services. It is well reported that privatization (or outsourcing) may sometimes undermine the quality of public services due to private operators' opportunism (e.g., Van Slyke, 2006). Indeed, private partners' opportunistic behavior may still be a problem in the crowdfunding of public services. For instance, the creator may embezzle the collected contributions. However, an advantage of public sector crowdfunding is that it allows crowdfunders to provide regular feedback to the operators by voting online with their dollars, which makes operators' opportunism less likely than in typical cases of privatization.[5]

Further, the crowdfunding projects are funded by *voluntary* contributions, which differs from taxpayer-funded government programs. Previous research has demonstrated that taxation can undermine the productivity of our economy by distorting the behaviors of taxpayers (Sandmo, 1976); for instance, higher tax rates cause workers to work less and entrepreneurs to invest less. However, there is no theoretical reason to believe that such a decrease in productivity exists in the case of crowdfunding. Therefore, with the quality of the services being equal, crowdfunding is preferred to tax-funded government operations. For instance, if school administrators use crowdfunding to purchase school equipment in times of austerity, there is no reason to oppose it as students are better off without making anyone worse off (and also without reducing anyone's productivity). Of course, to make the crowdfunding a success, school administrators should make a persuasive case that the school equipment will greatly improve the quality of school service.

---

[5] In many cases, public sector projects take several years to complete, and may involve several rounds of crowdfunding.

Another concern may be related to the government's steering capacity; governments can choose certain social problems while discriminating against others. In public sector crowdfunding, however, the financial resources come from the crowd rather than from the government. Both the social problems identified by the government and solutions proposed by the NPOs should face public evaluations on the platform of whether there exists sufficiently high demand (i.e., needs) from the citizenry. Indeed, public sector crowdfunding may reduce, not reinforce, governments' discretionary power of discriminating against certain groups of social members.

Nevertheless, we do not argue that public sector crowdfunding is a panacea. One potential problem of this innovation is that the crowdfunding platform may employ a selection mechanism that prefers a certain type of project over another. That is, there may be inherent biases built into the crowdfunding process. For instance, previous studies have consistently demonstrated that polarized, distinct, and extreme ideas tend to attract more attention on many Internet-based platforms (Hong 2013; Hong & Kim, 2016; Kim & Hong, 2015; Sunstein, 2018). This evidence may indicate that, if all else being equal, politically contentious ideas may attract more contributions on the platform than moderate ones. Further, the government's reliance on crowdfunding may also create equity concerns with regard to the revenue-raising capabilities of different communities. Previous research has found some evidence that the use of local option sales taxes (special-purpose taxes levied at the local level) exacerbated existing fiscal disparity across localities (Afonso, 2016; Zhao & Hou, 2008). Similar concerns may emerge regarding the government's use of crowdfunding if the fundraising is significantly more successful in wealthier communities than in less affluent counterparts. That is, the government's use of crowdfunding may introduce a new source of fiscal inequality across localities.

A potentially more critical problem may be that public sector crowdfunding involves a high level of uncertainty in terms of future contributions from the crowd. It thus cannot be used to fund public services that require sustainable financial support over a long period. One possible solution may be that governments make a binding commitment to fund the residual amounts (i.e., the amounts that were not successfully raised from the crowd). However, such government commitment may crowd out voluntary contributions from the public, undermining the utility of the innovation. Further, philanthropic crowdfunding may not be successful if used to fund projects that require a large amount of funding. For instance, the average funding goal of the projects we analyzed was about 2,500 USD per project, which is relatively small. Without any provision of substantive rewards to funders, it is unclear whether public sector crowdfunding can be applied to fund long-term projects requiring a sizeable amount of contributions.

## 8. CONCLUSION

This research explored whether and to what extent collaborative governance between public and private sectors could improve the outcomes of crowd-based financing. Among many forms of collaborative governance, we focused on the partnerships between government agencies and NPOs in the context of philanthropic crowdfunding. Such a public-private partnership in crowdfunding may be potentially beneficial to both parties; from the viewpoint of philanthropy, NPOs may collect a greater amount of donations using crowdfunding, whereas, from the viewpoint of government agencies, they may achieve public missions with substantially lower spending, which could free governments that are under fiscal distress of funding social programs.

The availability of crowdfunding as a mode of financing could certainly present an opportunity for private sector organizations, but the performance may be limited because of information asymmetry between the creators and funders. This issue may be especially pronounced in philanthropic crowdfunding, as NPOs should collect contributions from the broad public (i.e., the crowd) who has little knowledge about their business and operations. Due to information asymmetry, the creator may suffer moral hazards, and the funders may be unwilling to donate, anticipating the creator's moral hazard.

In this study, we explored whether and to what extent the partnerships between government agencies and NPOs could help to mitigate the information asymmetry problem, thereby improving crowdfunding performance. Our sample includes philanthropic crowdfunding projects created by NPOs for achieving public missions. We examined whether the projects with government support achieved a greater success rate or collected a greater amount of funding than comparable projects without any government involvement. Our findings indicate that the existence of government support is associated with a 64% increase in crowdfunding success rate and a 55% increase in the crowdfunding amount.

We then hypothesized that government involvement provides some type of accreditation or certification, attesting that the projects aim to achieve public rather than private goals, thereby improving citizen trust and mitigating the information asymmetry problem. The fact that government participation or support alleviates information asymmetry between creators and funders may indicate that it substitutes the creator's efforts toward project transparency. This research tests this "substitution hypothesis" between government support and transparency. We find that the positive impact of government support on crowdfunding success rates or funding amounts is significantly greater when the transparency level of the project is relatively low. This finding is suggestive but compelling evidence that the positive impact of government involvement stems primarily from alleviating citizens' concern of moral hazard and from promoting their trust in the projects.

The combined evidence of this study shows the potential that governments may actively take advantage of crowdfunding to achieve collaborative innovation and public goals. This collaborative financing may be especially beneficial for governments under fiscal distress; public missions may be pursued with voluntary contributions by the citizenry rather than through costly taxation. In other words, governments under fiscal distress may take advantage of this crowd-based financing to "co-fund" public projects with the citizenry. Moreover, governments may also use crowdfunding to gauge the collective needs for the project under consideration; crowdfunding allows governments to observe the aggregate amount of what citizens are willing to pay to bring the project into reality, and citizens can "vote with their dollars" online. These benefits may, of course, accrue in any charitable crowdfunding, even without any government involvement. However, our evidence shows that government involvement may certainly help realize these benefits by alleviating the information asymmetry problem.

**Figure 1.** Theoretical Relationship among Government Support, Transparency, and Crowdfunding Performance

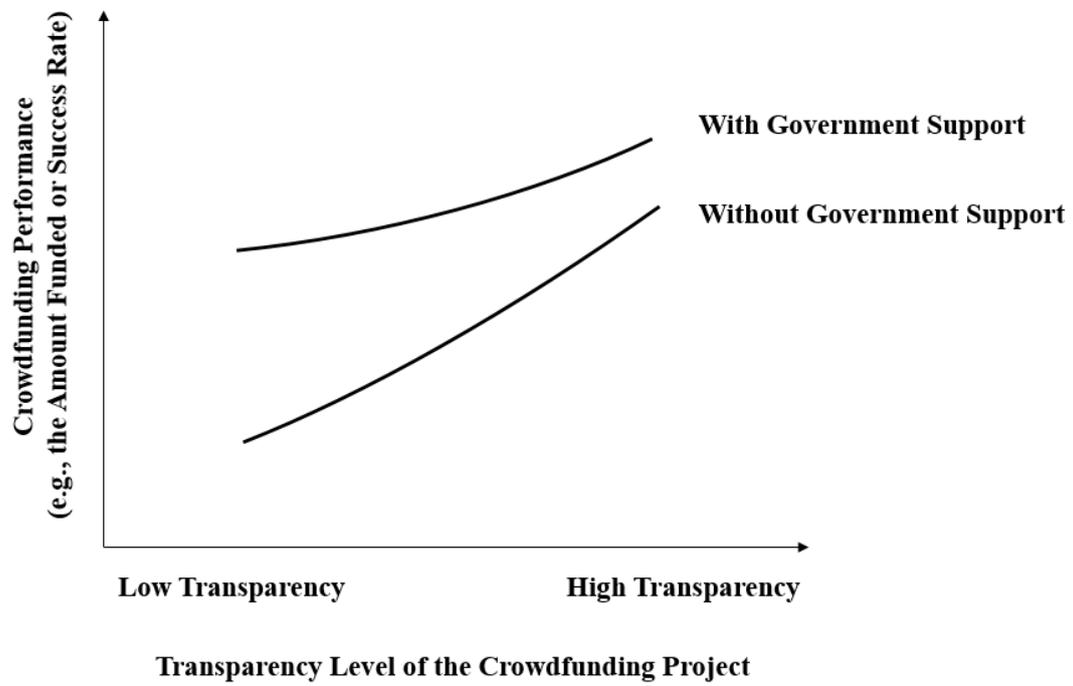

**Figure 2.** Empirical Relationship among Government Support, Transparency, and Crowdfunding Performance

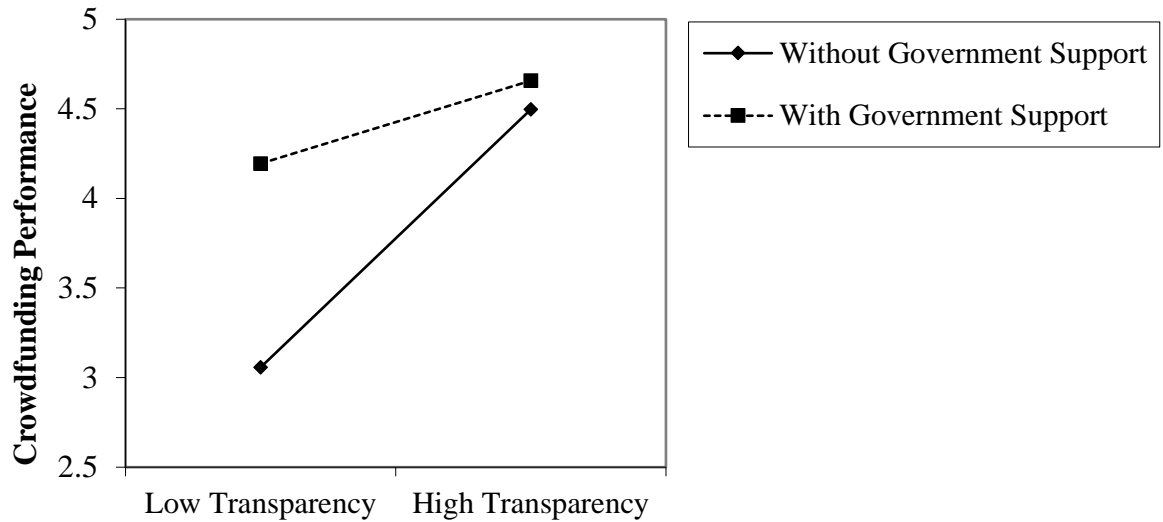

Note: The values in this figure are based on the coefficients in column 4 of Table 2, and were calculated directly from the coefficients. To do so, we set covariates at their mean values, took the product of the coefficients and values, and then summed up these quantities (3.78). We then added this value to what we obtained for different configurations of the key variables (i.e., transparency, government support, and the interaction between the two). In the figure, "high transparency" and "low transparency" are defined as transparency variables, representing the mean plus one standard deviation and the mean minus one standard deviation, respectively.

**Table 1.** Summary Statistics

| Variable | Mean | Standard Deviation | Min. | Max. |
|---|---|---|---|---|
| Success rate (*in log*) | 3.976 | 1.41 | 0 | 6.915 |
| Funding amount (*in log*) | 13.88 | 1.40 | 9.210 | 16.09 |
| Transparency | 0.000 | 0.695 | -1.008 | 3.566 |
| Government support | 0.336 | 0.475 | 0 | 1 |
| Duration (*in log*) | 3.359 | 0.340 | 2.398 | 4.663 |
| Keep-it-all type | 0.182 | 0.387 | 0 | 1 |
| Funding goal (*in log*) | 14.51 | 0.704 | 13.12 | 16.81 |

**Table 2**. Government Support and the Performance of Crowdfunding

|  | *Dependent Variable*: Crowdfunding Success Rate (*in log*) | | | |
| --- | --- | --- | --- | --- |
|  | (1) | (2) | (3) | (4) |
| Government Support (0 or 1) | 0.748** (0.305) |  | 0.637** (0.292) | 0.649** (0.285) |
| Transparency |  | 0.644** (0.143) | 0.596** (0.150) | 1.036** (0.299) |
| Transparency x Government Support (0 or 1) |  |  |  | -0.703** (0.335) |
| Duration (*in log*) | 0.156 (0.461) | 0.076 (0.467) | -0.017 (0.457) | -0.135 (0.454) |
| Keep-it-all Type (0 or 1) | -0.550 (0.348) | -0.892** (0.317) | -0.668* (0.339) | -0.716** (0.341) |
| Constant | 3.256* (1.642) | 3.497** (1.683) | 3.519** (1.638) | 4.218** (1.604) |
| N | 110 | 110 | 110 | 110 |
| $R^2$ | 0.150 | 0.191 | 0.226 | 0.250 |

Note: Standard errors in parentheses, * $p < 0.10$, ** $p < 0.05$; all models include a set of dummy variables for various types of beneficiaries of the project.

**Table 3**. Government Support and the Performance of Crowdfunding

| | *Dependent Variable*: The Amount of Crowdfunded Contributions (*in log*) | | | |
|---|---|---|---|---|
| | (1) | (2) | (3) | (4) |
| Government Support (0 or 1) | 0.683** | | 0.550* | 0.563* |
| | (0.308) | | (0.297) | (0.291) |
| Transparency | | 0.699** | 0.653** | 1.072** |
| | | (0.134) | (0.142) | (0.268) |
| Transparency x Government Support (0 or 1) | | | | -0.672** |
| | | | | (0.319) |
| Duration (*in log*) | 0.060 | -0.077 | -0.146 | -0.257 |
| | (0.446) | (0.444) | (0.438) | (0.436) |
| Keep-it-all Type (0 or 1) | -0.407 | -0.690** | -0.511 | -0.560* |
| | (0.339) | (0.310) | (0.324) | (0.325) |
| Funding Goal (*in log*) | 0.562** | 0.447** | 0.486** | 0.496** |
| | (0.205) | (0.193) | (0.199) | (0.203) |
| Constant | 5.478 | 7.541** | 6.943** | 7.462** |
| | (3.303) | (3.135) | (3.243) | (3.235) |
| $N$ | 110 | 110 | 110 | 110 |
| $R^2$ | 0.173 | 0.238 | 0.265 | 0.287 |

Note: Standard errors in parentheses, * $p < 0.10$, ** $p < 0.05$; all models include a set of dummy variables for various types of beneficiaries of the project.

**Acknowledgments:**

The authors would like to thank the anonymous reviewers and editors for their truly constructive comments. This work was supported by the Ministry of Education and the National Research Foundation of the Republic of Korea (NRF-2017S1A3A2067636).